\documentclass[aps,prl,onecolumn,superscriptaddress]{revtex4}  

\usepackage{xcolor}
\usepackage{amsmath,bm}
\usepackage{graphicx}
\usepackage{upgreek}

\setcitestyle{super}

\newcommand{\vecJmax}{\mathbf{J_{\mathrm{max}}}}
\newcommand{\Gdark}{\Gamma_{\mathrm{dark}}}
\newcommand{\Gpump}{\Gamma_{\mathrm{p}}}
\newcommand{\Gprobe}{\Gamma_{\mathrm{pr}}}
\newcommand{\DB}{\Delta B}

\newcommand{\pTrtHz}{\mathrm{pT}/\sqrt{\mathrm{Hz}}}
\newcommand{\fTrtHz}{\mathrm{fT}/\sqrt{\mathrm{Hz}}}

\newcommand{\lr}[1]{\left(  #1 \right)}

\newcommand{\vecB}{\mathbf{B}}
\newcommand{\vecJ}{\mathbf{J}}

\begin{document}

\title{Magnetocardiography on an isolated animal heart \\with a room-temperature optically pumped magnetometer}

\author{Kasper Jensen}
\thanks{corresponding author}
\affiliation{Niels Bohr Institute, University of Copenhagen, Blegdamsvej 17, 2100 Copenhagen, Denmark}
\author{Mark Alexander Skarsfeldt}
\affiliation{Department of Biomedical Sciences, Faculty of Health and Medical Sciences, University of Copenhagen, Blegdamsvej 3, 2200 Copenhagen N, Denmark}
\author{Hans C. St{\ae}rkind}
\affiliation{Niels Bohr Institute, University of Copenhagen, Blegdamsvej 17, 2100 Copenhagen, Denmark}
\author{Jens Arnbak}
\affiliation{Niels Bohr Institute, University of Copenhagen, Blegdamsvej 17, 2100 Copenhagen, Denmark}
\author{Mikhail~V.~Balabas}
\affiliation{Niels Bohr Institute, University of Copenhagen, Blegdamsvej 17, 2100 Copenhagen, Denmark}
\affiliation{Department of Physics, St Petersburg State University, Universitetskii pr. 28, 198504 Staryi Peterhof, Russia}
\author{S{\o}ren-Peter Olesen}
\affiliation{Department of Biomedical Sciences, Faculty of Health and Medical Sciences, University of Copenhagen, Blegdamsvej 3, 2200 Copenhagen N, Denmark}
\author{Bo Hjorth Bentzen}
\affiliation{Department of Biomedical Sciences, Faculty of Health and Medical Sciences, University of Copenhagen, Blegdamsvej 3, 2200 Copenhagen N, Denmark}
\author{Eugene S. Polzik}
\affiliation{Niels Bohr Institute, University of Copenhagen, Blegdamsvej 17, 2100 Copenhagen, Denmark}

\maketitle

%%%%%%%%%%%%%%%%%%%%%%%%%%%%%%%%%%%%%%%%%%%%%%%%%%%%%%%
%\newpage

\section{Abstract} 

\textbf{
Optically pumped magnetometers are becoming a promising alternative to cryogenically-cooled superconducting magnetometers for detecting and imaging  biomagnetic fields. Magnetic field detection is a completely non-invasive method, which allows one to study the function of excitable human organs with a sensor placed outside the human body. For instance, magnetometers can be used to detect brain activity or to study the activity of the heart. 
We have developed a highly sensitive miniature optically pumped magnetometer based on cesium atomic vapor kept in a paraffin-coated glass container. The magnetometer is optimized for detection of biological signals and has high temporal and spatial resolution. It is operated at room- or human body temperature and can be placed in contact with or at a mm-distance from a biological object. 
With this magnetometer, we detected the heartbeat of an isolated guinea-pig heart, which is an animal widely used in biomedical studies. 
In our recordings of the magnetocardiogram, we can in real-time observe the P-wave, QRS-complex and T-wave associated with the cardiac cycle.
We also demonstrate that our device is capable of measuring the cardiac electrographic intervals, such as the RR- and QT-interval, and detecting drug-induced prolongation of the QT-interval, which is important for medical diagnostics.
}

\section{Introduction}
The cardiac function  is based on conduction of electrical signals, which activate the heart muscle so blood can be pumped throughout the body. Electrodes placed on the body surface can record this cardiac electrical activity. This method is called electrocardiography (ECG) and is commonly used for diagnosing heart diseases. The electrical signals propagating through the heart generate magnetic fields, which can be detected and imaged with sensitive magnetometers placed outside the body. This method called magnetocardiography (MCG) can be advantageous to use in certain cases including when measuring the electrical activity in the fetal heart 
as the method is better than ECG to discern the signal of the fetus from that of the mother.
Detection of the magnetic field from the heart was first done using pick-up coils \cite{Baule1963} and later with a superconducting quantum interference device (SQUID) magnetometer, which together with the development of a shielded room lead to improved signal-to-noise ratio of the observed magnetocardiograms \cite{Cohen1970apl}.
SQUID magnetometers are highly sensitive but have the major drawback that they require cryogenic cooling. 
Optically pumped magnetometers  (OPMs) can achieve similar high sensitivity but they do not require cryogenics and are considered promising for many applications, in particular for detecting brain activity\cite{Xia2006,Boto2017neuroimage,Boto2018nature} (magnetoencephalography) and for magnetocardiography on the adult \cite{Bison2009apl,Lembke2014} and fetal heart \cite{Wyllie2012ol,Alem2015PhysMedBio,Eswaran2017}.
OPMs based on absorption of light and operating in the so-called spin-exchange relaxation-free (SERF) 
regime\cite{Kominis2003nature} are now commercially available \cite{quspin}. SERF magnetometers have some disadvantages, which include an operating temperature of several hundreds of degrees Celsius, which leads to the need for thermal insulation, the fact that they should be operated in zero magnetic field, and their limited bandwidth and dynamic range.
Our OPM is based on cesium atomic vapor and is operated at room- or human body temperature. 
This allows the magnetometer to be placed in contact with or at a mm-distance from biological tissue. 
Our type of cesium magnetometer can be used to detect time-varying signals with frequency components ranging from DC to MHz and it can have a large and tunable bandwidth.
The magnetometer has achieved sub-fT/$\sqrt{\mathrm{Hz}}$ sensitivity when used for detecting radio-frequency magnetic fields\cite{Wasilewski2010prl}.
More recently, the magnetometer was employed in biomedical experiments where it was used to detect action potentials from an isolated animal nerve \cite{Jensen2016scirep}.

Early diagnosis and intervention on cardiac conduction abnormalities 
before birth can be lifesaving. Although the development and improvement of echocardiography has been instrumental in managing fetal arrhythmia, it is limited as it measures the mechanical consequence of the arrhythmia rather than the electrical activity underlying the arrhythmia. Fetal MCG with detection of the cardiac electrical activity can help to understand the electrical mechanism underlying fetal arrhythmias\cite{Leeuwen1999} such as, bradycardia, atrio-ventricular block\cite{Zhao2008} and supraventricular tachycardia\cite{Wakai2003circ} and hence guide correct treatment. Another example of this is the fetal diagnosis of long~QT syndrome\cite{Cuneo2013,Cuneo2016}. 
Congenital long~QT syndrome is a heritable arrhythmia syndrome, with a prevalence of 1:2500 in healthy live births\cite{Schwartz2009}, characterized by delayed repolarization of the ventricular cardiac action potential, which manifests as a prolonged QT-interval on the ECG. It is most often caused by mutations in genes responsible for the repolarization of the ventricular cardiac action potential, and results in an increased risk of ventricular arrhythmia, and consequently is an important contributor to sudden cardiac death in newborns and young\cite{Lieve2015}. 
Detection of long~QT syndrome in fetuses is important in order to decrease fetal morbidity\cite{Cuneo2015}. However, because fetal ECG recordings are currently not feasible and echocardiography cannot be used to detect prolonged QT-intervals, there is a need for novel non-invasive tools for prenatal long QT syndrome detection.
Non-invasive magnetocardiography would enable healthcare professionals  to diagnose these abnormalities. 
Using the guinea pig heart that has comparable size (see Methods section), heart rate\cite{Steinburg2013,ACOG2009} and electrical properties\cite{Farraj2011toxic} to that of heart from a human fetus at gestational age of 18-22 weeks, we demonstrate that our magnetometer is capable of detecting the cardiogram in real-time and capable of detecting changes in the QT interval of the cardiogram. This shows that OPMs are promising devices for biomedical studies and for diagnosis of fetal heart diseases.

\section{Results}
\subsection{Guinea-pig electrocardiogram}
\begin{figure*}[th]
\centering
\includegraphics[width=1\textwidth]{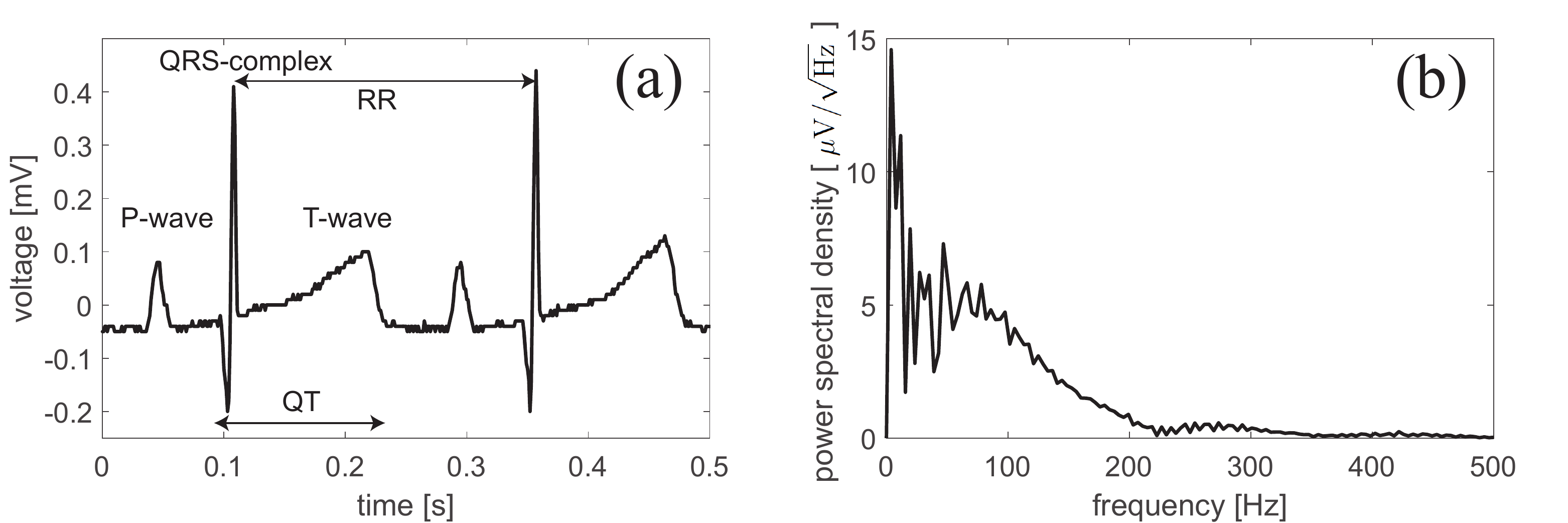}
\caption{Electrical recording on an isolated guinea-pig heart placed in a Langendorff setup. (a) Electrocardiogram showing the P-wave, the  QRS-complex and the T-wave. Electrographic intervals RR and QT are depicted with arrows. (b) Power spectral density of the electrocardiogram of a single heart beat lasting approximately 0.25~s ($\mathrm{RR}=247$~ms corresponding to 243~bpm.)}
\label{fig:ECG}
\end{figure*}

In this study we perform experiments on isolated guinea-pig hearts
 having about the size of a human fetus heart. 
A detailed description of this preparation is provided in the Methods section. Before any magnetic field measurements are initiated, the
electrically active, spontaneously pumping heart being perfused 
with saline 
is first placed in a separate Langendorff setup\cite{Bell2011cardiology}. An example ECG measured in the Langendorff setup is shown in Fig.~1(a). The P-wave, QRS-complex and the T-wave are all clearly visible. A single heart beat [corresponding to the time-interval from 0 to 0.25~s in Fig.~\ref{fig:ECG}(a)] has frequency components mainly in the range from DC-200~Hz as seen in Fig.~\ref{fig:ECG}(b). 
As discussed in detail below, our magnetometer is capable of detecting such signals with high sensitivity.

\subsection{Experimental setup}
Our magnetometer is based on cesium atomic vapor kept in a paraffin-coated miniature glass container [see Fig.~\ref{fig:heartcellsetup}(a)] with inner size 
$5~\textrm{mm} \times 5~\textrm{mm} \times 5~\textrm{mm} $ and  $0.7~\textrm{mm}$ glass-wall thickness. 
The  vapor cell is kept at room- or human body temperature and is placed inside a multi-layer cylindrical magnetic shield. 
Circularly polarized pump (wavelength $\approx 895$~nm) and repump (wavelength $\approx 852$~nm) laser light on resonance with the cesium D1 and D2 transitions, respectively, is used for optical pumping [Fig.~2(c,d)]. 
Linearly polarized laser light (wavelength $\approx 852$~nm) near-resonant with the cesium D2 transition ($\approx 1600$~GHz detuning) is used to probe the spin-polarized cesium atoms.
Any magnetic field will affect the atomic spins  which in turn will lead to a rotation of the probe light polarization. This is detected with a balanced detection scheme [see Fig.~2(c)] which gives a magnetometer signal from which we can infer the magnetic field (see details below).

\subsection{Model of the magnetometer}
The cesium atoms are described by their total angular momentum 
 $\vecJ = \lr{J_x,J_y,J_z}$ which in the absence of any magnetic field is optically pumped along the x-direction using circular polarized light. A fully spin-polarized ensemble of cesium atoms corresponds to the spin vector $\vecJmax=\lr{4 N_A,0,0}$ for $N_A$ atoms in the $F=4, m_x=4$ ground state hyperfine sublevel. 
In a simplified model of the magnetometer, the time-evolution of the spin vector can be described by the following differential equation \cite{Ledbetter2008pra}
\begin{equation}
\frac{d \vecJ }{dt}  = \gamma \vecJ \times \vecB + \Gpump \vecJmax -\lr{\Gpump+\Gprobe + \Gdark}\vecJ ,
\label{eq:dJdt}
\end{equation}
where $\vecB$ is the external magnetic field and $ \gamma=2.2\cdot 10^{10}~\mathrm{rad/s}$ is the cesium gyromagnetic ratio. 
In the above, optical pumping happens with a rate $\Gpump$ and the spin decays with a total rate $\Gamma=\Gpump+\Gprobe+\Gdark$, where $\Gprobe$ is the probe-induced decay rate and $\Gdark$ is the decay rate in the absence of any light. $\Gpump$ and $\Gprobe$ are proportional to the power of the pump and probe laser light, respectively.
The atomic spin and thereby the magnetic field is detected by sending a near-resonant and linearly-polarized probe laser beam trough the atoms and recording the light polarization rotation which is proportional to the spin-component $J_z$ along the probe direction\cite{Jensen2016scirep}.

If we assume that the magnetic field is static or varying slowly on the time-scale set by $T_2 = 1/\Gamma$, we can solve Eq.~(\ref{eq:dJdt}) in the steady-state, and if we 
furthermore assume that the magnetic field is small $|B|^2 \ll \DB^2$, where $\DB = \Gamma/\gamma$, we find that the magnetometer signal 
$S(t)\propto J_z(t) = J_0 B_y(t)/\DB$, where $J_0=J_{\mathrm{max}} \Gpump/\Gamma$  (see Methods section). I.e., the magnetometer signal is proportional to the $y$-component of the magnetic field.
On the other hand, if the field is varying on a time-scale comparable to $T_2$ or faster, we find that the magnetometer signal is
\begin{equation}
S(t)  = c \int_{t'=-\infty}^{t'=t} 
e^{-\Gamma \lr{t-t'}}  B_y(t') dt',
\label{eq:SignalConv}
\end{equation}
where $c$ is a normalization constant and we again assumed that the magnetic field is small (see Methods section). 
In this case, the magnetometer signal is a convolution of the $y$-component of the magnetic field with the  magnetometer response function $c \cdot e^{-\Gamma t}$.
The magnetic field $B_y(t)$ can then be obtained by deconvolving the magnetometer signal $S(t)$ with the response function. 

\subsection{Calibrating the magnetometer}
The magnetometer is calibrated before any biomagnetic measurements. Even though the magnetometer is placed inside a magnetic shield, there will be a small residual static field. Three sets of coils are used to apply magnetic fields along the $x$-, $y$-, and $z$-directions in order to cancel the residual field such that the total field is approximately zero.
In order to measure the temporal profile $e^{-\Gamma t}$ of the magnetometer response function, an extra coil is used to apply a short square pulse (100~$\mu$s duration) of magnetic field along the $y$-direction. 
The observed magnetometer signal [solid line in Fig.~\ref{fig:Signals}(a)] increases rapidly and then decays exponentially with time-constant $T_2=4.0$~ms and is in good agreement with the model given by Eq.(\ref{eq:dJdt}). 
Notice that due to power broadening by the probe, pump and repump light, the decay-time is significantly shorter than the coherence time in the dark  $T_2^{\mathrm{dark}}=1/\Gdark > 50~\mathrm{ms}$. For our type of vapor cell, a long $T_2^{\mathrm{dark}}$ is possible due to the paraffin-coating inside the vapor cell which preserve the  spin of the moving cesium atoms during collisions with the glass walls. 

We then apply a magnetic field pulse with the temporal shape of a single sinusoidal oscillation with 10~ms period and independently calibrated amplitude. 
By numerically deconvolving the measured signal [dashed line in Fig.~\ref{fig:Signals}(a)] with the un-normalized response function $e^{-\Gamma t}$ and comparing the amplitude of the deconvoluted signal with the independently calibrated amplitude, we can extract the normalization constant $c$, and calculate the magnetic fields in nT or pT as shown  Fig.~\ref{fig:Signals}(b). 

We also record the magnetometer signal when there is no magnetic field pulse applied. The noise spectrum of the magnetometer signal is shown in 
Fig.~\ref{fig:Signals}(c). At frequencies above 1~kHz, the noise is mainly due to the shot noise of the probe light. At lower frequencies the noise level is higher (see Discussion section). The noise spectrum in magnetic field units found by deconvolution is shown in Fig.~\ref{fig:Signals}(d). We achieve the best sensitivity of around 120~fT/$\sqrt{\mathrm{Hz}}$ for frequencies in the 100-500~Hz range, and around or better than 300~fT/$\sqrt{\mathrm{Hz}}$ for frequencies in the wider range 5-1000~Hz.

We note that the \textit{atomic bandwidth} $1/\left( 2\pi T_2\right)$ can be widely tuned by adjusting the laser powers. For our specific settings, the atomic bandwidth is 40~Hz. The advantage of the deconvolution procedure is that it allows one to correctly find the temporal profile of magnetic fields with much higher frequency components up to the one half times the data-acquisition sample-rate which in our case is $\frac{1}{2}\times 10 \; \mathrm{kHz} =5 \; \mathrm{kHz}$. 
The magnetic field from the heart is expected to have frequency components in the DC-200~Hz range which is outside the atomic bandwidth. The deconvolution procedure is therefore necessary in order to correctly find the temporal profile of a magnetocardiogram, and in particular for finding the correct amplitudes of the QRS-complex and the P-wave which are the fastest-varying features of the cardiogram.
The deconvolution procedure also amplifies high frequency noise, and in case the signal is small, additional low-pass or bandpass filtering may be required in order to obtain sufficient signal-to-noise ratio. 
When recording the magnetic field from the heart, we choose to low-pass filter the deconvoluted signals using a Type~I Chebyshev filter with a cut-off frequency of either 250~Hz or 500~Hz. This \textit{magnetometer bandwidth} of 250~Hz or 500~Hz is high enough to capture the temporal profiles and amplitudes of the QRS-complex and P-wave and low enough to supress high-frequency noise and thereby improve the signal-to-noise ratio.

\subsection{Magnetic field from the isolated guinea-pig heart}
\begin{figure*}[th]
\centering
\includegraphics[width=0.9\textwidth]{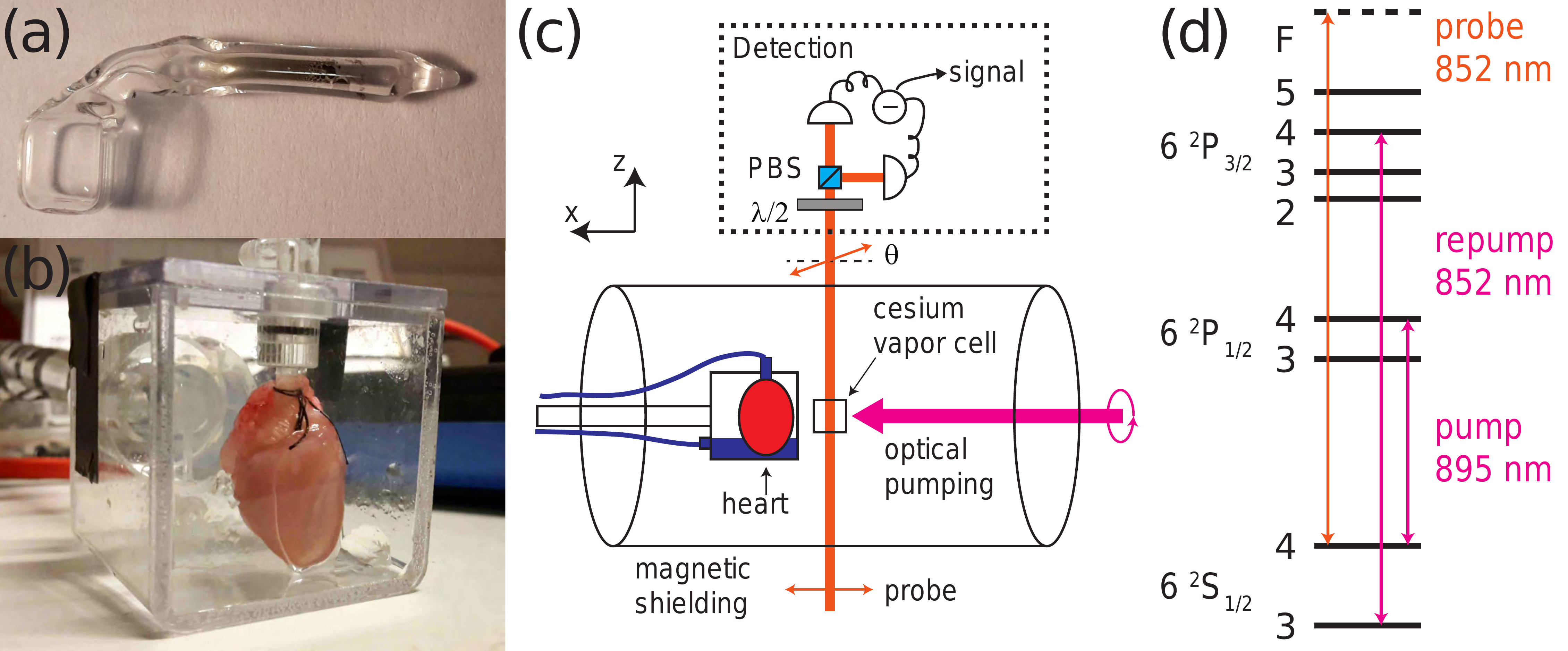}
\caption{Experimental setup.
(a) Picture of the cesium vapor cell. (b) Picture of an isolated guinea-pig heart inside a plastic chamber. (c) Schematics of the experimental setup. $\lambda/2$: half-wave plate, PBS: polarizing beamsplitter. 
The magnetic field from the heart affects the cesium atomic spins which in turn rotates the polarization of the probe light by an angle $\theta$. This polarization rotation is detected with a balanced polarimeter (shown inside the dashed-line box). (d) Cesium atom level scheme and laser frequencies.}
\label{fig:heartcellsetup}
\end{figure*}

\begin{figure*}[th]
\centering
\includegraphics[width=\textwidth]{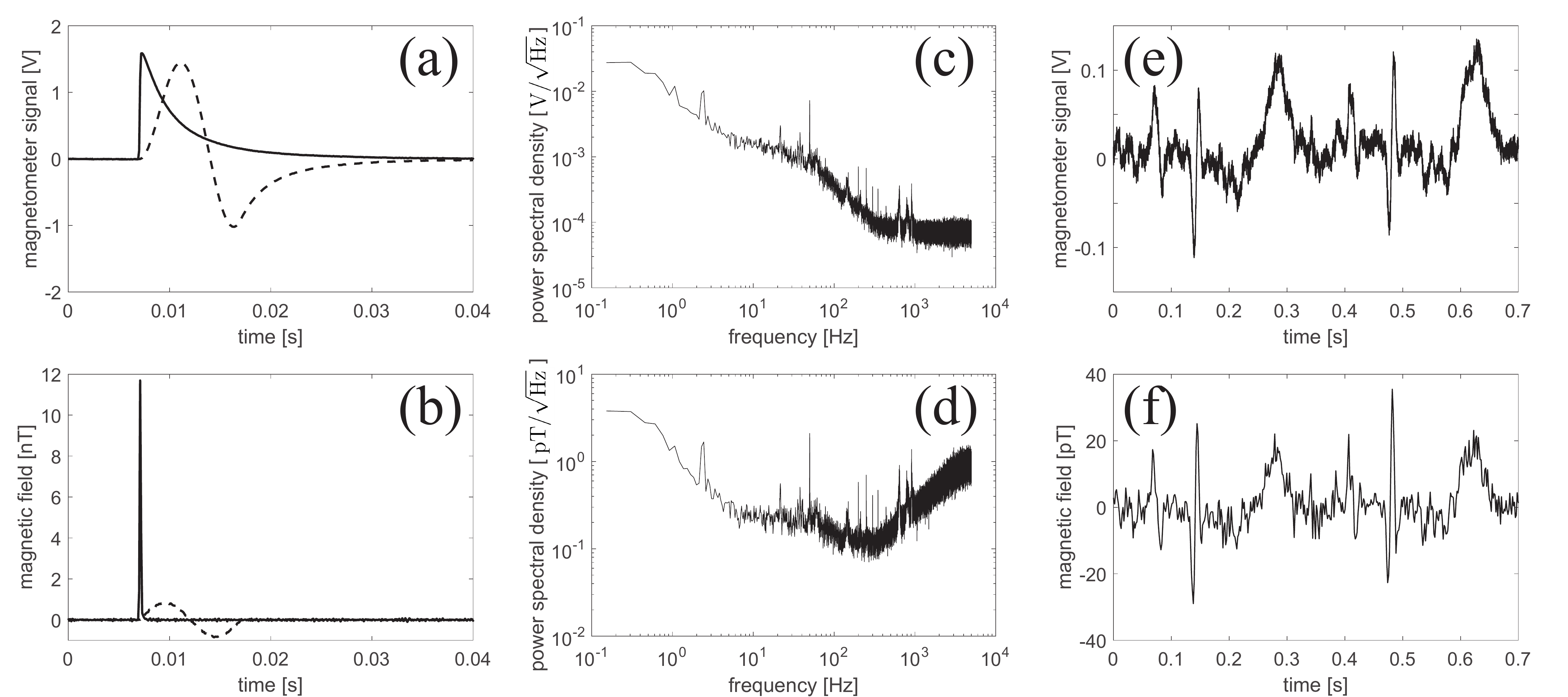}
\caption{Magnetic field measurements.
(a) Magnetometer signal in volts. Solid line: 100~$\mu$s square magnetic pulse applied. Dashed line: single sine wave magnetic pulse with $10$~ms period applied. Signals have been averaged 10~times. 
(b) Magnetic field in nT as calculated by deconvolving the signals in (a). 
(c) Noise spectrum of the magnetometer signal in V/$\sqrt{\mathrm{Hz}}$ without any applied magnetic field. Noise spectrum has been averaged 10~times. (d) Noise spectrum in $\pTrtHz$. 
(e) Example of a magnetometer signal (without averaging) in volts when a guinea-pig heart is placed close to the magnetometer. 
(f) Magnetic field from the heart in pT. The deconvoluted signal was low-pass filtered using a Type I Chebyshev filter with a cut-off frequency of 250~Hz to supress high-frequency noise.}
\label{fig:Signals}
\end{figure*}

During the biomagnetic measurements, the spontaneously beating guinea-pig heart is put in a watertight acrylic container  [see Fig.~\ref{fig:heartcellsetup}(b)] and placed close to the vapor cell. The position of the heart relative to the vapor cell can be set using a three-axis translation stage with travel range of 25~mm. The smallest possible distance between the heart and the center of the vapor cell is 5.2~mm (2~mm thickness of acrylic container, 2.5~mm inside radius of vapor cell and 0.7~mm glass-wall thickness). Our magnetometer measures the $y$-component of the magnetic field (see Fig.~\ref{fig:heartcellsetup}c) which is a component of the field parallel to the surface of the heart.

An example of a real-time recording of the magnetometer signal is shown in Fig.~\ref{fig:Signals}(e).  We note that it is possible to see the P-wave, QRS-complex and T-wave in real-time without the use of averaging. The magnetic field found by deconvolution is shown in Fig.~\ref{fig:Signals}(f). The QRS complex has duration $\approx 25$~ms and is varying on a time-scale comparable to $T_2$. The deconvolution procedure is therefore necessary to correctly find the QRS amplitude which for this particular magnetocardiogram is approximately $55$~pT peak-to-peak.

The actual shape or morphology of the magnetocardiogram depends on the position of the heart relative to the magnetometer. To illustrate this, we measured MCGs at $4\times 4 = 16$ different relative positions covering a 15~mm $\times$ 15~mm area close to the heart  (see Supplementary Figure S1). The relative position was changed by translating the plastic chamber containing the heart in the $y$-$z$-plane   
(see Fig.~\ref{fig:heartcellsetup}c) in steps of 5~mm.
In Supplementary Figure~S2, we clearly see that the temporal shape of the magnetocardiograms depend on the position. 
This is even more apparent in the zoom-ins of the QRS-complexes shown in Supplementary Figure~S3.
For instance, it seems like the magnetic field changes direction/sign depending on whether the magnetometer is in front of the right or left side of the heart.
Also the amplitude of the QRS-complex depend on position. In those measurement, the peak-to-peak amplitude varied in the range 50-120~pT.
We note that because we only use a single magnetic field sensor, and not a whole magnetometer array, measurements at different positions are recorded at different times. 
This is a problem if the heart is not beating at the same frequency over the whole scan period, something which is challenging to achieve because the heart is isolated from the guinea-pig, spontaneously beating and is sensitive to temperature and flow/pressure fluctuations of the Krebs-Henseleit buffer (see Methods section) used to perfuse the heart. Despite such possible issues, our magnetometer can indeed be placed close to the heart and be used to measure the magnetic field around the heart with high (here 5~mm) spatial resolution and good signal-to-noise ratio.

\begin{figure*}[th]
\centering
\includegraphics[width=0.33\textwidth]{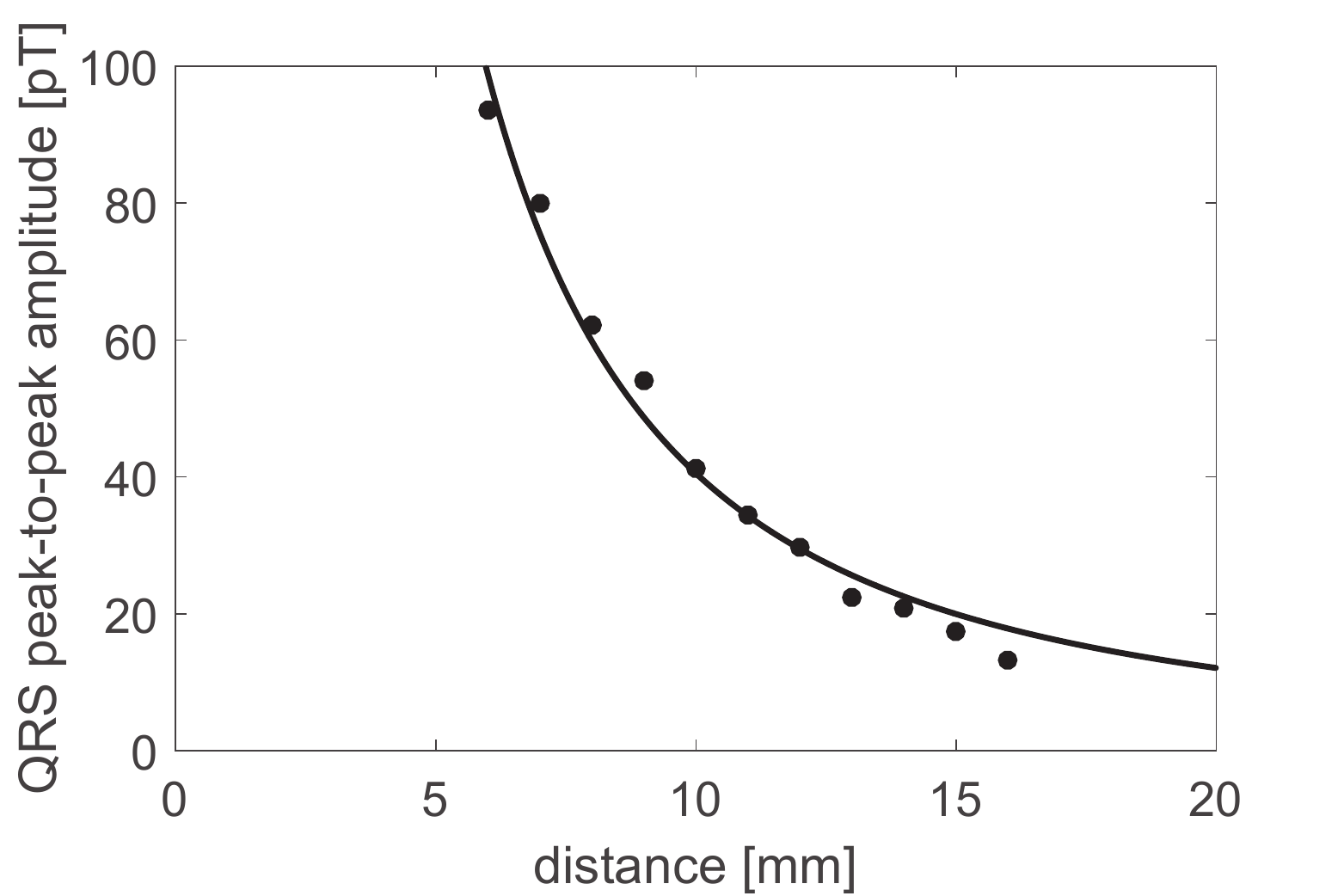}
\caption{Peak-to-peak value of the QRS complex as a function of distance between the heart and magnetometer.}
\label{fig:varydist}
\end{figure*}

For one of the positions in the $y$-$z$-plane, we also varied the distance between the heart and the magnetometer from 6-16~mm (along the $x$-direction, see Fig.~\ref{fig:heartcellsetup}c). In those measurements, the QRS amplitude dropped from $\approx 100 \; \mathrm{pT}$  peak-to-peak to  less than 15~pT peak-to-peak as seen in Fig.~\ref{fig:varydist}. The QRS amplitude as a function of distance was fitted to a function $B_{\mathrm{pp}}^{\mathrm{QRS}} \propto x^{-n}$, where the fitted value $n=1.7(2)$ was obtained. 
The fit can be used to estimate the signal at larger distances relevant for fetal magnetocardiography. For instance, for $x=50$~mm, the extrapolation of the fitted curve gives $B_{\mathrm{pp}}^{\mathrm{QRS}}(50 \;\mathrm{mm})=2.4 \;\mathrm{pT}$ which is $\approx 40$ times smaller than the field at $x=6 \; \mathrm{mm}$.

\subsection{Drug-induced prolongation of the QT interval}
\begin{figure*}[th]
\centering
\includegraphics[width=\textwidth]{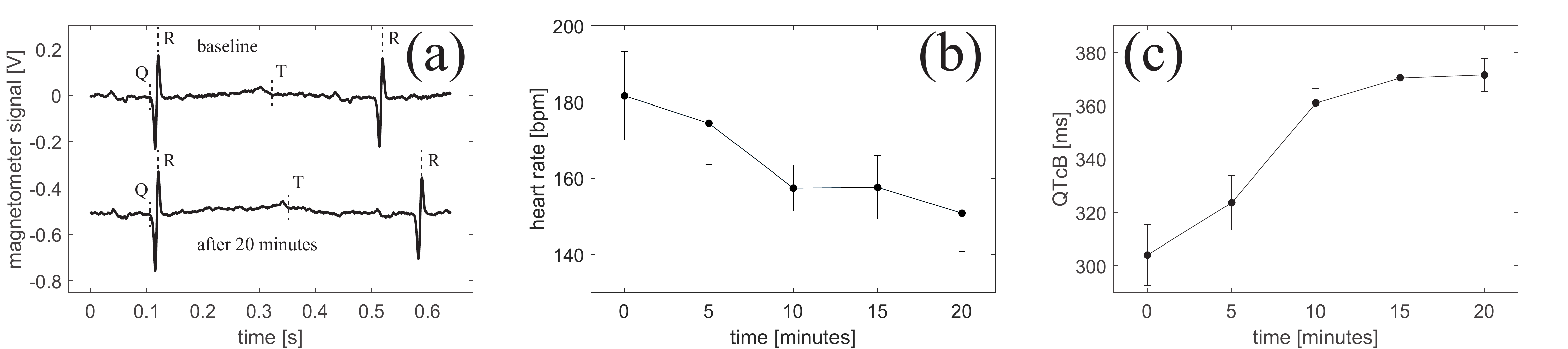}
\caption{Effects of I$_{\mathrm{Kr}}$ inhibition on heart rate and QT-interval. 
(a) Example magnetometer signals recorded before and after administration of 0.06 $\mu$M E-4031. Traces are the averages of approximately 10 heartbeats.
(b) Heart rate in beats per minute: bpm = 60/RR-interval [s].
Points are mean values averaged over 5 animals. Error bars represent the standard error of the mean.
(c) QT-interval corrected for heart rate (Bazett correction\cite{Hamlin2003toxic}).
}
\label{fig:RRandQT}
\end{figure*}

We pharmacologically mimicked inherited long QT syndrome using a potent ion channel blocker (E-4031), that in sub-micro molar concentrations blocks the rapid delayed rectifying repolarizing current (I$_{\mathrm{Kr}}$), and results in prolongation of the QT-interval (see Methods section).
Before the measurements, the isolated hearts were allowed to stabilize in the watertight container for approximately 30 minutes. Then, a baseline measurement of the MCG was performed before adding QT-prolonging drug (E-4031) to the perfusion fluid. 
MCGs were measured every 5th minute for a 20-minute time frame [see example MCGs in Fig.~\ref{fig:RRandQT}(a)]. 
Throughout the experiment, the coronary artery flow was kept constant by adjusting the perfusion pressure in order to ensure a stable physiological temperature of the heart. 
We repeated the experiment using in total five isolated guinea-pig hearts.
As seen in Fig.~\ref{fig:RRandQT}(b,c), we were able to detect the QT-prolongation of 0.06 $\mu$M E-4031 and detect a decrease in heart rate using our optically pumped magnetometer.

%%%%%%%%%%%%%%%%%%%%%%%%%%%%%%%%%%%%%%%%%%%%%%%%%%%%%%%%%%%%%%%%%%%%%%%%%%%%%%%%%%%%%%%%%%%%%%%%%%%%%%%%%%%%%%%%%%%%%
\section{Discussion}

In conclusion, we have demonstrated that our optically pumped magnetometer can be used for real-time monitoring of the heartbeat of an isolated guinea-pig heart. We have furthermore imaged the magnetic field around the heart, measured how the field decays with distance, and showed that the magnetometer can be used in biomedical studies and to diagnose cardiac 
repolarization abnormalities such as the long QT syndrome. 
The measurements are possible due to the high sensitivity of the magnetometer and its operating temperature and small size which makes it possible to place the magnetometer in contact with or at just a few mm distance from a biological object.
The current sensitivity of the magnetometer is around 120-300~$\fTrtHz$ for a wide frequency range 5-1000~Hz. 
The sensitivity is ultimately limited by the quantum spin-projection noise of the cesium atoms which is estimated to be  
4~$\fTrtHz$ (see discussion of the sensitivity in the Methods section).
Achieving this sensitivity would lead to an improved signal-to-noise ratio of a factor 30-75 which would make the quality of the recorded MCGs outstanding.
Such an improved sensitivity would also allow for real-time recordings of MCGs at larger distances. Based on our measurements on the guinea-pig heart, we conclude that 
real-time detection of the heartbeat of a human fetus at gestational age of 18-22~weeks,
where the heart-sensor distance is estimated to be $\geq 5$~cm, 
should be possible.
We also note, that our OPM works at physiological temperatures, which would allow for probes designed for use inside the human body. The advantage with that approach is that signals would be much larger when the sensor is close to the object under investigation.

Further development of an array of OPMs\cite{Borna2017physmedbio,Alem2017oe} operating at physiological temperature would be beneficial as this would allow for fast and simultaneous recording of the field at many positions.
This would enable 3D-reconstruction of the distribution of ionic currents flowing inside the heart, similar to what is done in neuroscience when a SQUID magnetometer array is used to image brain activity.
Note however, that the cardiac intervals such as RR and QT do not depend on the position of the magnetometer with respect to the heart. This means that for diagnosing elevated heart rate or QT-prolongation, only a single sensor is required.
In fetal-magnetocardiography, using a few sensors will be beneficial: one sensor would record the fetal-MCG, another sensor the mother’s MCG, and a few other sensors would monitor and correct for drifts in the background field.
An array of OPMs working at physiological temperatures may also be interesting for biomedical or pre-clinical studies. It would allow for monitoring the propagation of action potentials with high spatial resolution in for instance isolated nerves\cite{Jensen2016scirep} or brain tissue. The combination of an OPM array and a guinea-pig heart could also be interesting for imaging and studying the local electrical conductivity of the heart\cite{Marmugi2016scirep}.
In conclusion, we believe that OPMs are promising devices for biomedical and pre-clinical studies, and that OPMs will have clinical use as a new diagnostic tool that will allow clinicians to do better diagnostics and prenatal care.

%%%%%%%%%%%%%%%%%%%%%%%%%%%%%%%%%%%%%%%%%%%%%%%%%%%%%%%%%%%%%%%%%%%%%%%%%%%%%%%%%%%%%%%%%%%%%%%%%%%%%%%%%%%%%%%%%%%%%

\section{Methods}

\subsection{Evolution of the spin vector in a magnetic field}
The evolution of the cesium atomic spins in an external magnetic field is modeled using Eq.~(\ref{eq:dJdt}).
If we assume that the magnetic field is static or varying slowly on the time-scale set by $1/\Gamma$, we can solve the equation in the steady-state and find the solution
\begin{eqnarray}
J_x^{ss} &=& J_0 
\frac{B_x^2+\left( \DB \right)^2}{|B|^2 + \left( \DB \right)^2 }, \\
J_y^{ss} &=& J_0 
\frac{B_x B_y - B_z \DB}{|B|^2 + \left( \DB \right)^2 } ,\\
J_z^{ss} &=& J_0 
\frac{B_x B_z + B_y \DB}{|B|^2 + \left( \DB \right)^2 } ,
\end{eqnarray}
where $|B|^2=B_x^2+B_y^2+B_z^2$, $\DB = \Gamma /\gamma$ and $J_0 = J_{\mathrm{max}} \Gpump/ \Gamma$.
If we  furthermore assume that the magnetic field is small $|B|^2 \ll \DB^2$, we find the steady state spin vector $ \mathbf{J}^{ss} \approx J_0 \left(1, -B_z/\DB,B_y/\DB \right)$. 

We can also solve Eq.~(\ref{eq:dJdt}) for the case where the magnetic field is varying on a time-scale comparable to $1/\Gamma$ or faster.
We will assume that the field can be written as 
$\mathbf{B}(t) = B_0 \mathbf{\hat{x}} + \delta \mathbf{B}(t)$, 
where $B_0$ is a static field pointing in the $x$-direction
and $\mathbf{\delta  B}(t)=\left( \delta B_x(t) , \delta B_y(t) ,\delta B_z(t) \right)$ is a small time-varying magnetic field which can be treated as a perturbation. To first order in $\mathbf{\delta  B}(t)$, we find the magnetometer signal
\begin{equation}
S(t) \propto J_z(t) = 
\gamma J_0 \int_{t'=-\infty}^{t'=t} e^{-\Gamma \lr{t-t'}} 
\left\{  \cos \left[2 \pi \nu_L \lr{t-t'}\right]  \delta B_y(t') 
- \sin \left[2 \pi \nu_L \lr{t-t'}\right]  \delta B_z(t') 
\right\}  dt',
\label{eq:S(t)conv}
\end{equation} 
where $\nu_L=\gamma  B_0 /\lr{2 \pi}$ is the Larmor frequency in Hz.
We see that the magnetometer signal is a convolution of $ \delta B_y(t)$ with the magnetometer response function 
$e^{-\Gamma t} \cos \left[2 \pi \nu_L t\right]$ plus the convolution of $ \delta B_z(t)$ with the response function
$-e^{-\Gamma t} \sin \left[2 \pi \nu_L t\right]$. I.e., the magnetometer will mainly be sensitive to magnetic fields which are transverse to the static field and which have frequency components that are within a certain bandwidth set by $\Gamma$ around the Larmor frequency $\nu_L$. Both $\nu_L$ and $\Gamma$ are widely tunable such that the magnetometer can be optimized for detection of specific time-varying magnetic fields. 
In our previous work we used non-zero static $B_0$-fields for detection of radio-frequency magnetic fields \cite{Wasilewski2010prl} and magnetic fields from animal nerve impulses \cite{Jensen2016scirep}. Here, we instead choose a static field close to zero as the heartbeat has frequency components mainly in the range from DC-200~Hz. For $B_0=0$, Eq.~(\ref{eq:S(t)conv}) simplifies to the expression
\begin{equation}
S(t) \propto J_z(t) = \gamma J_0 \int_{t'=-\infty}^{t'=t} 
e^{-\Gamma \lr{t-t'}}  \delta B_y(t') dt'.
\end{equation}
We see that for $B_0=0$, the magnetometer is only sensitive to the $y$-component of the magnetic field, as opposed to the case $B_0 \neq 0$ where the magnetometer is sensitive to both the $y$- and $z$-components of the field.

\subsection{Discussion of magnetometer sensitivity}

The sensitivity of our magnetometer is around 120~fT/$\sqrt{\mathrm{Hz}}$ for frequencies in the 100-500~Hz range, and around or better than 300~fT/$\sqrt{\mathrm{Hz}}$ for frequencies in the wider range 5-1000~Hz [see Fig.~\ref{fig:Signals}(d)].
Improving the sensitivity will be beneficial for  biomedical experiments. Our current magnetometer is based on free-space laser beams with $\approx 1$~m propagation distances, and for that reason the setup is quite sensitive to vibrations and air-flows. A more compact magnetometer is expected to be crucial for improving the sensitivity. 50~Hz noise and possibly low-frequency (DC-200~Hz) magnetic field noise is also problematic. Active stabilization of the background magnetic field would improve those noise sources. 
The sensitivity is ultimately limited by the quantum spin-projection noise of the cesium atoms\cite{Jensen2016scirep} which at room-temperature is estimated to be
$1/\left( \gamma \sqrt{T_2 2 n V}\right) =
 8$~$\fTrtHz$ using the density of atoms $n=3.0\cdot 10^{16}$ (for $T\approx 20^\circ$C), sensing volume $V=\left(5 \: \mathrm{mm} \right)^3$ and $T_2=4$~ms. At human-body temperature 37$^\circ$C, the density of atoms is four times higher leading to a quantum-limited sensitivity of 4~$\fTrtHz$. Our type of magnetometer has previously achieved sub-femtotesla sensitivity mainly limited by quantum noise for detection of radio-frequency magnetic fields \cite{Wasilewski2010prl}.

\subsection{Isolated guinea pig heart preparations}
The experiments were approved by the Danish Veterinary and Food Administration, Ministry of Environment and Food of Denmark under License No. 2014-15-2934-01061 and were carried out in accordance with the European Community Guidelines for the Care and Use of Experimental Animals.
The experiments were performed using female Dunkin Hartley guinea pigs (300-450 g, Charles River, Saint-Germain-Nuelle, France). Guinea pigs were anesthetized with an intraperitoneal injection of pentobarbital 200 mg/ml and lidocaine hydrochloride 20 mg/ml (Glostrup Apotek, Denmark), dose 0.150 ml/100 g BW. A tracheostomy was performed and the guinea pigs were ventilated (3.5 ml/60 strokes/min) through a rodent ventilator (7025 Rodent ventilator, Ugo Basile, Italy). The hearts were cannulated \textit{in situ} through an incision of the aorta and connected to the Langendorff retrograde perfusion set-up (Hugo Sachs Elektronik, Harvard Apparatus GmbH, March, Germany). The hearts were retrogradely perfused at a constant perfusion pressure of 60 mmHg with a 5$^\circ$C, pH 7.8, cardioplegia buffer (in mM: NaCl 110, NaHCO$_3$ 25, KCl 16, CaCl$_2$ 1.2, MgCl$_2$ 16). 
At cardiac arrest the hearts were submerged in 5$^\circ$C cardioplegia buffer and transported to a different
laboratory where the magnetometer was located.
Here, the hearts were revived by a retrogradely perfusion of 37$^\circ$C Krebs-Henseleit buffer (in mM: NaCl 120, NaHCO$_3$ 25, KCl 4, MgSO$_4$ 0.6, NaH$_4$PO$_4$ 0.6, CaCl$_2$ 2.5, Glucose 11) saturated with 95\% O$_2$ and 5\% CO$_2$. The hearts were continuously perfused with a coronary flow ranging from 9-14~ml/min. The spontaneously beating hearts were mounted in a custom made watertight acrylic container (inner size 35~mm, wall thickness 2~mm) during the following measurements. The aortic perfusion pressure was determined with an ISOTEC transducer (Hugo Sachs Elektronik), and the coronary flow was measured with an ultrasonic flowmeter (Transonic Systems INC, USA). Both were connected to an amplifier (Hugo Sachs Elektronik). Perfusion pressure and coronary flow signals were sampled at a frequency of 2 k/s and converted by a 16/30 data acquisition system from PowerLab systems (ADInstruments, Oxford, UK) and monitored by using LabChart 7 software (ADInstruments).
\subsection{Pharmacology}
The hearts were perfused with 0.06 $\mu$M E-4031 (Sigma-Aldrich, Darmstadt, Germany), a class III anti-arrhythmic drug, that blocks the repolarizing I$_{\mathrm{Kr}}$/hERG/K${_\mathrm{v}}$11.1 current, and thereby prolong the ventricular action potential demonstrated by prolonged QT-intervals. 

\subsection{Size of guinea-pig hearts}
Six isolated guinea-pig hearts were used for the measurements presented here. The size of the hearts are given in Supplementary Table~S1. The mean biventricular diameter was 16.8~mm and the mean length was 23.3~mm. 
The size of the hearts are similar to that of a human fetus at gestational age of 18-22~weeks\cite{Xinyan2015}.

% Requirement from Scientific Reports
% https://www.nature.com/srep/journal-policies/editorial-policies#availability
%\subsection{Data Availability}
%The datasets generated during and/or analysed during the current study are available from the corresponding author on reasonable request.

%\bibliography{BIBQuantop4} 
%\bibliographystyle{unsrt} 

%\bibliographystyle{naturemag} 

%%%%%%%%%%%%%%%%%%%%%%%%%%%%%%%%%%%%%%%%%%%%%%%%%%%%%%%%%%%%%%%%%%%%%%%%%%%%%%%%%

%%%%%%%%%%%%%%%%%%%%%%%%%%%%%%%%%%%%%%%%%%%%%%%%%%%%%%%%%%%%%%%%%%%%%%%%%%%%%%%%%

\section*{Acknowledgements}
This work was supported by 
the Danish Quantum Innovation Center (QUBIZ)/Innovation Fund Denmark, 
the ERC grant QUMAG, 
and the U. S. Army Research Laboratory and the U. S. Army Research Office under contract/grant number Grant No. W911NF-11-0235.

\section*{Author Contributions}
K.J., M.A.S, H.C.S, J.A. B.H.B performed the experiments. 
M.A.S and B.H.B. prepared the animal hearts.
M.V.B. designed and fabricated the vapor cell. 
K.J., M.A.S, and B.H.B. wrote the manuscript.
All authors discussed the results and commented on the manuscript.
B.H.B., S.-P.O. and E.S.P supervised the research.

%\section*{Additional Information}
%\textbf{Competing financial interests:} The authors declare no competing financial interests.

%%%%%%%%%%%%%%%%%%%%%%%%%%%%%%%%%%%%%%%%%%%

%\newpage

\section{Figures}

%\newpage

%%%%%%%%%%%%%%  	supplementary figures %%%%%%%%%%%%%%%%%%%%%%%%%%%%%%
\clearpage
\section{Supplementary Information}

\begin{figure*}[th]
\centering
\includegraphics[width=0.5\textwidth]{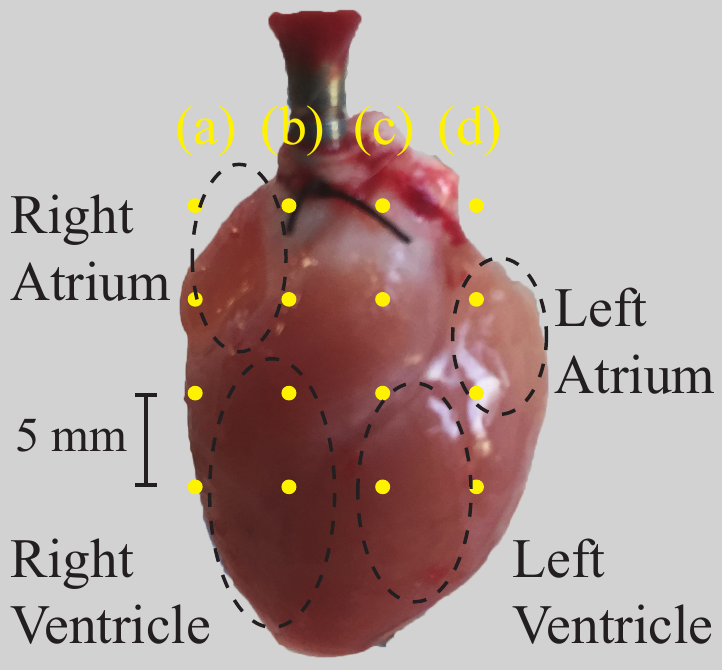}
\end{figure*}

\textbf{Supplementary Figure~S1:} 
Picture of an isolated guinea-pig heart. 
Approximate locations of Right Atrium,  Right Ventricle, Left Atrium, and Left Ventricle are indicated in the figure with dashed ellipses.
In the magnetic field measurements shown later in Supplementary Figures~2 and 3, the heart was translated in the $y$-$z$-plane such that different regions of the heart were positioned in front of the cesium vapor cell. The approximate relative positions (a), (b), (c), (d), ... are indicated in the figure with yellow dots.

\clearpage

\begin{figure*}[th]
\centering
\includegraphics[width=\textwidth]{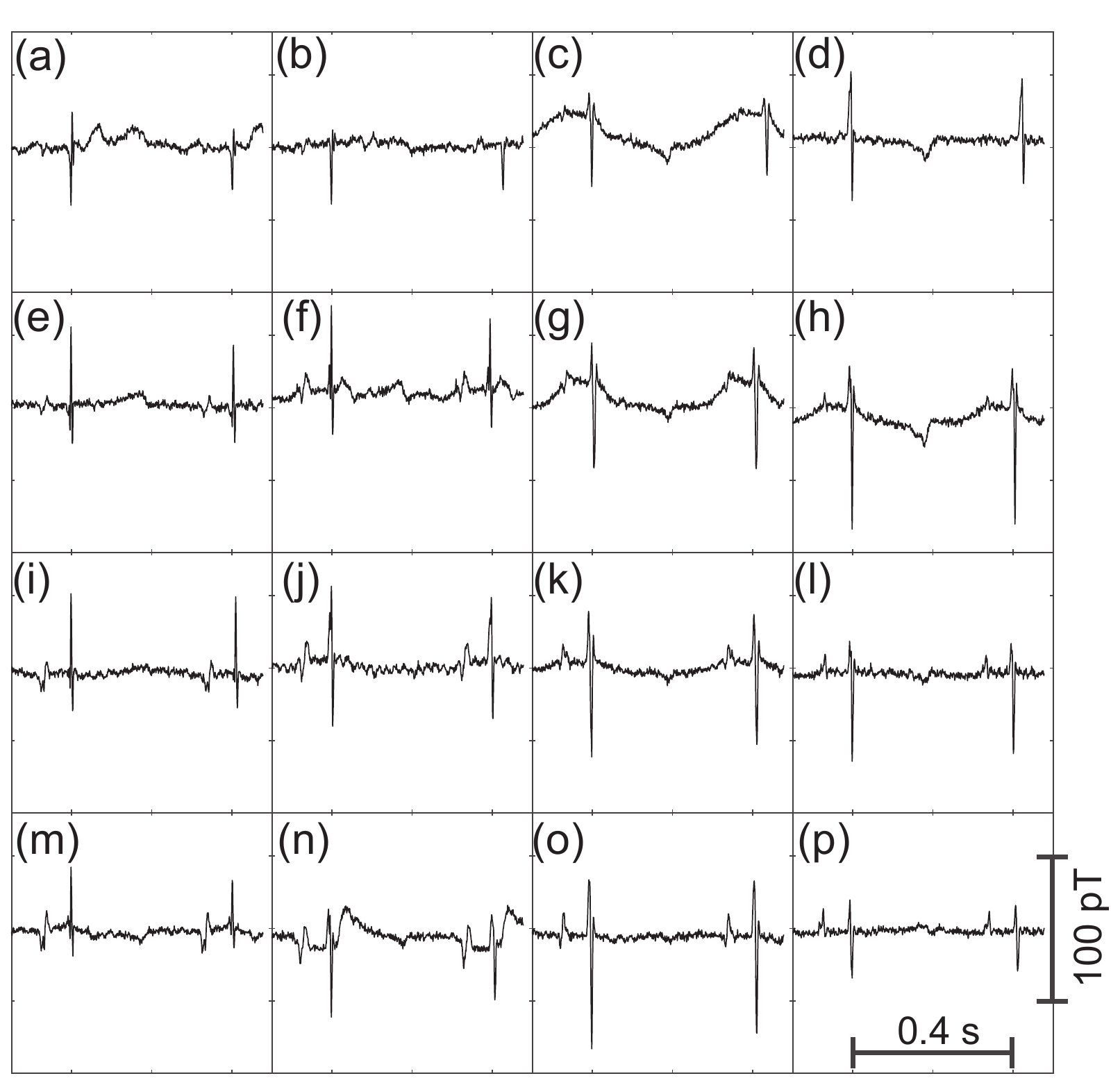}
\end{figure*}

\textbf{Supplementary Figure~S2:} Magnetic field from the heart recorded at 16 different positions  of the heart relative to the magnetometer (yellow dots in Supplementary Figure~S1). 
In these recordings, the signals were converted to magnetic field units by deconvolution and low-pass filtering using a Type I Chebyshev filter with a cut-off frequency of 500~Hz.
For each position, we recorded the magnetic field for 4~seconds. During those 4 seconds, the heart beat approximately 10 times. In the data analysis, the 10 heartbeats were averaged in order to improve the signal-to-noise ratio.

\clearpage

\begin{figure*}[th]
\centering
\includegraphics[width=\textwidth]{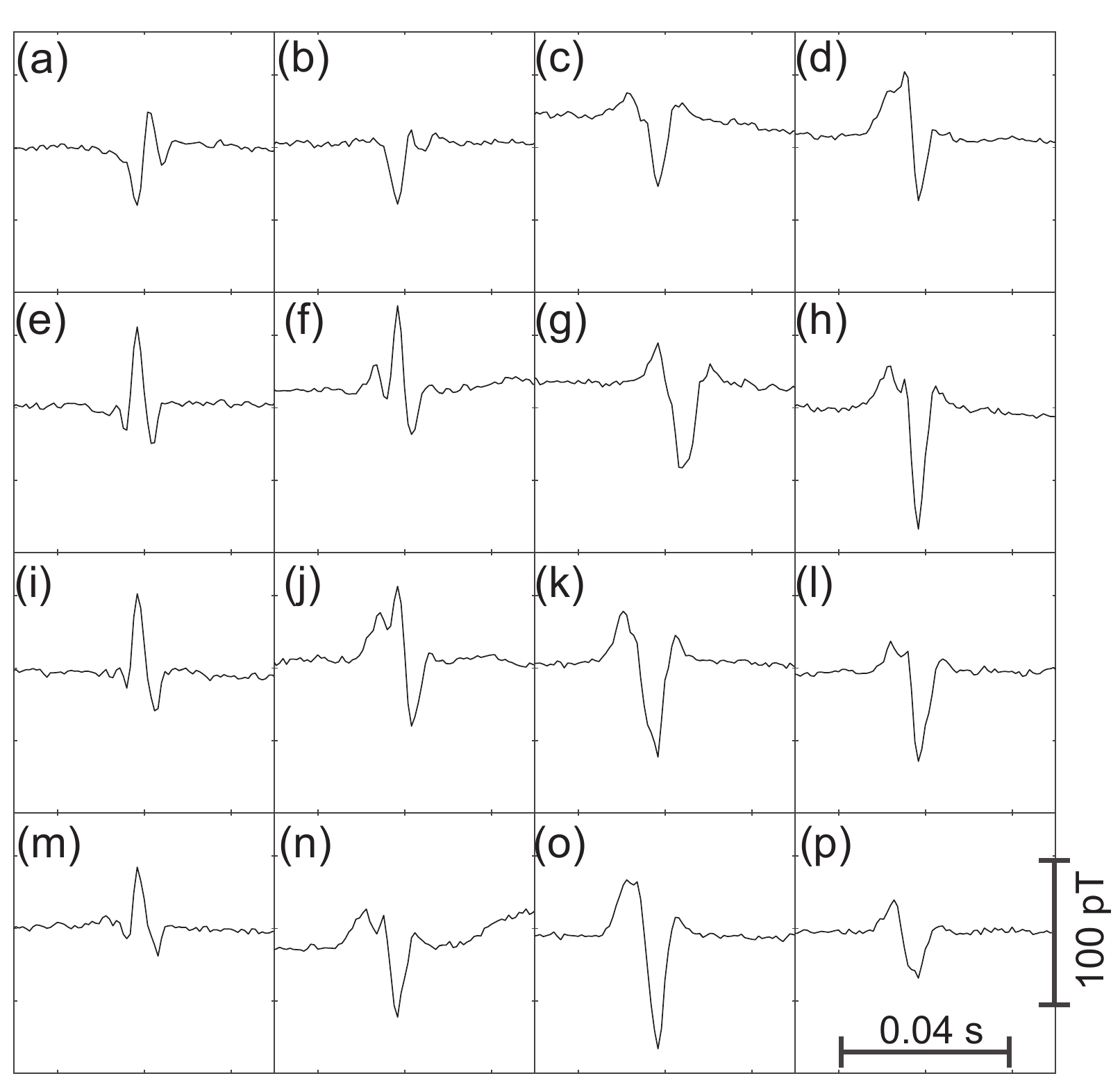}
\end{figure*}

\textbf{Supplementary Figure~S3:} Zoom-ins on the QRS-complexes of the magnetocardiograms shown in Supplementary Figure~S2.
Notice that the morphology and the peak-to-peak amplitude of the QRS-complex depend on the relative position between the heart and the vapor cell.

\clearpage

\begin{table}

	\centering
		\begin{tabular}{ l | c | c| c| c| c |c | c| c}
			heart & 1 & 2 & 3 & 4 & 5 & 6 & mean & std\\
			\hline
			biventricular diameter [mm] & 18 & 14 & 14 & 18 & 17 & 20 & \textbf{16.8} &2.4\\
			length [mm] & 23 & 23 & 24 & 22 & 23 & 25 & \textbf{23.3} &1.0\\
		\end{tabular}
%		\label{tab:size}
\end{table}

\textbf{Supplementary Table~S1:} Measured size of six isolated guinea-pig hearts.

%%%%%%%%%%%%%%%%%%%%%%%%%%%%%%%%%%%%%%%%%%%%%%%%
\end{document}